\documentstyle[aps,preprint,epsf]{revtex}

\begin{document}

\draft

\title{Negative superfluid densities in superconducting films in a parallel
magnetic field}

\author{F.Zhou*, B.Spivak}

\address{Department of Physics, University of Washington,
Seattle, WA 98195, USA}
\date{\today}

\maketitle

\begin{abstract}

In this paper we develop a theory of mesoscopic fluctuations in 
disordered thin superconducting films in a parallel magnetic field.
At zero temperature and sufficiently strong magnetic field the system 
undergoes a phase transition into a state characterized by 
a superfluid density, which is random in sign. Consequently, in this
region, random supercurrents are spontaneously created in the ground state
of the system, and it belongs to the same universality class as the two 
dimensional $XY$ spin glass with a random sign of the exchange interaction.

\end{abstract}


Recent experiments on thin superconducting films in parallel magnetic field
$^{[1]}$ have rekindled interest in this field. 
If the thickness of the films is small enough,
the orbital effect of the magnetic field can be neglected and the 
suppression of 
superconductivity in the film is due to the Zeeman effect
$^{[2-4]}$.  
It has been observed that the resistance of such films at low 
temperatures and high enough  magnetic fields exhibits very
slow relaxation in time$^{[1]}$. This 
behavior is characteristic for spin and superconducting glasses.
Below we discuss a possibility that 
mesoscopic fluctuations of superconducting parameters in 
disordered films account for such a behavior.
Usually, in the limit where the electron elastic mean free path $l$
exceeds the Fermi wave length $\hbar/k_{F}$, mesoscopic fluctuations of
various physical parameters 
of superconductors are smaller than their
averages$^{[5-8]}$. 
Thus, they hardly affect macroscopic observable quantities.
However, there are situations where mesoscopic fluctuations determine 
macroscopic properties of superconducting samples. One example is
a superconductor in a magnetic field close to the upper critical field
$H_{c2}$,
 where the magnetic field dependence of the 
superconducting critical temperature is determined by the mesoscopic 
fluctuations$^{[9]}$.
In this paper we consider the case, where the magnetic field is parallel
to the thin superconducting film and the main contribution to the 
suppression of superconductivity by the magnetic filed is due to Zeeman 
splitting of electron spin energy levels. 
We will show that at low temperatures $T$ and high enough magnetic fields $H$, 
parallel to the film, the 
system exhibits a transition into a state where local superfluid density 
$N_{s}(\vec{r})$ (which is the ratio between the supercurrent density
 $\vec{J_{s}}$ 
and the superfluid velocity $\vec{V_s}$) has random sign. In this case
the 
system belongs to the same universality class as the two-dimensional 
$XY$
spin glass model.
The idea that the superfluid density can be negative has a long history 
$^{[5-7, 10-13]}$. However, in the absence of magnetic 
fields and at zero temperature in 
slightly disordered superconductors ($\xi_{0}\gg l\gg {\hbar}/{k_{F}}$)
the variance of 
the superfluid density, averaged over the superconducting coherence length
$\xi_{0}=\sqrt{{D}/{\Delta_0}}$, turns out to be much smaller than
its average$^{[5-8]}$  $\langle(\delta N_{s})^2\rangle =G^{-2}(\langle
N_{s}\rangle)^{2}\ll (\langle N_{s}\rangle)^{2}$,
where $\delta N_{s}=N_{s}-\langle
N_{s}\rangle$, $G$ is the dimensionless
conductance of the normal metal film, in
units
of ${e^2}/{\hbar}$.
Here $D={v_{F}l}/{3}$ is
the diffusion coefficient, $\Delta_{0}$ is the value of the order
parameter at $T,H=0$; $v_F$ is the Fermi velocity, and the 
brackets $\langle\rangle$ denote averaging over realizations of random
potential. 
This means, that as long as $k_{F}l\gg \hbar$, the regions where the 
superfluid
density is negative are rare and do not contribute 
significantly to macroscopic properties of superconductors.
The situation in the presence of a magnetic field parallel to the film is 
different, because the average superfluid density decays with $H$ faster 
than its variance.
Hence, at high enough magnetic field the amplitude of the mesoscopic 
fluctuations of $N_{s}(\vec{r})$ becomes larger than the average, and the
respective 
probabilities of having positive and negative signs of $N_{s}(\vec{r})$
are of the same order even at $k_F l \gg 1$. 

A theory of magnetic field induced phase transition, which does not 
take into account mesoscopic fluctuations predicts 
$^{[2,14,15]}$ that at low temperatures 
the superconductor-normal metal 
transition is of first or second order
depending on whether the parameter $\Delta_{0}\tau_{so}$ is larger or smaller 
than unity, respectively. Here $\tau_{so}$ is the spin-orbit relaxation
time.

Let us start with the case where $\Delta_{0}\tau_{so}\ll 1$.
At $T=0$ and within an approximation which neglects mesoscopic 
effects, the value of the critical magnetic field $H^{0}_{c}$ is the
result of the 
competition between the average superconducting condensation energy density
 $\langle E_{c}\rangle\sim \nu_0 \Delta_0^{2}$ and the 
polarization energy of the electron gas in the 
magnetic field.
Here $\nu_0$ is the average density of states in 
the metal on the Fermi surface. 
The average spin polarization energy
density of 
nonsuperconducting 
electron gas is of order of $\langle E_{p}(0)\rangle\sim \nu_{0}(\mu_{B}
H)^{2}$. Its 
relative change in the superconducting state is of order of 
$\langle E_{p}(0)\rangle -\langle E_{p}(\Delta)\rangle\sim
\frac{3}{4}\pi\Delta_0\tau_{so}\langle E_{p}(0)\rangle\ll 
\langle E_{p}(0)\rangle$$^{[16-18]}$.
 As a result we get an expression 
for the critical magnetic 
field $H^{0}_{c}=H_{cc}(\Delta_0\tau_{so})^{-\frac{1}{2}}\gg 
H_{cc}$. Here $H_{cc}={\Delta_0}/{\mu_B}$ is the
Chandrasekar-Clogston critical magnetic field of the superconductor-normal
metal transition for
$\Delta_{0}\tau_{so}\rightarrow\infty$ and $\mu_{B}$ is the Bohr
magneton.

Now let us consider the mesoscopic fluctuations of the quantities, 
discussed above, in a volume whose size is of the order of the coherence length 
$\xi_{0}$. The amplitude of mesoscopic 
fluctuations of the polarization energy is of order of 
$^{[19]}$ $\sqrt{\langle(\delta E_{p})^{2}\rangle}\sim
G^{-1}(\Delta_0\tau_{so}) 
\langle E_{p}(0)\rangle$, while its $\Delta$-dependent part is
of the order of
\begin{eqnarray}
\frac{\sqrt{\langle(\delta E_{p}(\Delta(H))-\delta E_{p}(\Delta
=0))^{2}\rangle}}{
\sqrt{(\langle \delta
E_{p}(0)\rangle^{2}}}
\sim 
\left (\frac{\Delta(H)}{\Delta_{0}}\right)^{2}. 
\end{eqnarray}
Here 
$\langle\Delta(H)\rangle=\Delta_{0}\sqrt{(H^0_{c}-H)/H^0_{c}}$ is the 
average superconducting order parameter. 
Since both the polarization energy and the condensation energy are 
fluctuating quantities, $\Delta(\vec{r})$ should also be
spatially fluctuating.
Let us consider a domain of size $L_D \gg \xi_{0}$ where the value of 
$\Delta(\vec{r})$ differs from its bulk value by a factor of order of
unity. 
An estimate for the energy of such a domain consists of three terms,
namely 
\begin{equation}
\frac{\delta E(\Delta)}{\nu_0 \Delta(H)^{2}dL_D^2} 
 \sim 
(C_1\frac{1}{G}\frac{\xi_0}{L_D}+
C_{2}\frac{H_c^0-H}{H^0_c}  
+C_{3}\frac{\xi_0^2}{L_D^2})
\end{equation}
where $d$ is the thickness of the film and $C_1,C_{2},C_{3}$ are factors
of order of unity. The first term in Eq.2 corresponds to
the $\Delta$-dependence of mesoscopic fluctuations of polarization
energy and has a random sign. When estimating this term we have taken into
account that domains of size $\xi_{0}$ make independent random sign
contributions 
into Eq.2. The second and third term are the average condensation
energy and surface (gradient) energy of the domain, respectively.
It follows from Eq.2 that if $L_D \sim
\xi(H)=\xi_{0}\sqrt{{H_c^0}/{|H-H_c^0|}}$ that there is an
interval of
magnetic fields near the critical one $H_c^0 -H \sim H^0_{c}/G^2$,
where the first term is larger than 
the second and the third ones. It means that, in this case the spatial 
distribution of 
$\Delta(\vec{r})$ is highly inhomogeneous and the amplitude of
the spatial fluctuations of $\Delta(\vec{r})$ is of order of its 
average, while the characteristic size of the domains is of order of 
$L_{D}\sim \xi(H=H_{c}^{0}(1-1/{G^2}))\sim \xi_{0}G$. 
Superfluid density in this region has a random sign as well. To see 
this, one should consider states with finite superfluid velocity 
$\vec{V}_{s}=(\nabla\chi + {2e}/{c}\vec{A})/m$, where
$\chi(\vec{r})$ is the
phase of the order parameter, 
$\vec{A}(\vec{r})$ is the vector
potential of a magnetic field perpendicular to the
film and $m$ is the electron mass. If $\vec{V}_{s}(\vec{r})$ is of the
order of the 
critical velocity, all three terms in Eq.2 are modified by factors of
order of
unity when compared with the case $\vec{V}_{s}=0$. 
The second and the third term in Eq.2 decrease with $\vec{V}_{s}$, while
the
first 
term is changed in a random direction. This 
means that at high enough magnetic fields, states with nonvanishing value
of
$\vec{V}_{s}(\vec{r})$ 
have lower energy than the states with $\vec{V}_{s}=0$, and that the
system is
unstable
 with respect to the creation 
of supercurrents of random directions. In this estimate we neglected the
energy of the magnetic field associated with $\vec{V}_{s}(\vec{r})$.
 Since at each point 
of the system the possible energy gain associated with finite value 
of $\vec{V}_{s}(\vec{r})$ is independent of the direction of
$\vec{V}_{s}$, the ground state of the system is highly degenerate and  
belongs to the same universality class as $XY$ spin glass with random sign 
of exchange interaction. 

It is important to mention that even in the case of small magnetic fields
in the presence of spin orbit scattering the time reversal symmetry is 
broken and the electron wave functions are complex. 
These currents flowing in the random directions
exist even in normal metals. By evaluating the diagrams shown in Fig.2a,
we derive the 
correlation function of the current 
density in normal metals($|\vec{r}-\vec{r'}|\gg {\hbar}/{k_F}$),
 
\begin{equation}
\langle J_{i}(\vec r) J_{j}(\vec r')\rangle\approx
\delta_{ij}\frac{e^{2}}{\hbar^4 d^{2}}\tau\tau_{so}(\mu_B
H)^{4}\delta(\vec r-\vec{r'}).
\end{equation}
Here $\tau={l}/{v_{F}}$ is the elastic
mean free time and $i,j$ are coordinate indexes.
It is important to note, however, that for a given configuration of the
scattering
potential
and at a given value of the external field the spatial distribution of
$\vec J(\vec r)$ is a unique function. This implies that the currents
described by Eq.3 do not
exhibit features which can be associated with superconducting glass
states. 

Below, we will be interested in supercurrents much larger than those
described by Eq.3. Such currents 
are spontaneously created at strong enough magnetic fields as a result of 
the instability associated with the random sign of superfluid density. 
Consider the Gorkov 
equation for $\Delta(\vec{r})$$^{[20]}$,

\begin{equation}
\Delta(\vec{r})=g \int d\vec{r'}K(\vec{r},\vec{r'};
H,\vec{A}(\vec{r}),\{\Delta(\vec r)\})
\Delta(\vec{r'}),
\end{equation}
where $K(\vec{r},\vec{r'}; H, \vec{A}(\vec r), \{\Delta(\vec r)\})=
kT \sum_{\epsilon}
 G^{\alpha \beta}_{\epsilon}(\vec{r},\vec{r'}; H, \vec{A}(\vec{r}),
\{\Delta(\vec r)\})
  \sigma^x_{\beta \mu}  \tilde{G}^{\mu\nu}_{-\epsilon}(\vec{r},\vec{r'};
H, \vec{A}(\vec{r}), 0) \\ \sigma^x_{\nu \alpha}$; 
$G^{\alpha \beta}_{\epsilon}(\vec{r}, \vec{r}'; H, \vec A(\vec r),
\{\Delta(\vec r)\})$ is
the exact one particle Matsubara Green function, $\alpha$,$\beta$, 
$\nu$, $\mu$
are spin indexes, $\sigma^{x}_{\alpha\beta}$ is the Pauli matrix and
$\epsilon=(2n+1)\pi kT$ is the Matsubara frequency. 
$g$ is the dimensionless interaction constant.
Both $\Delta(\vec{r})$ and $K(\vec{r},\vec{r'})$ in Eq.4 are
random functions of realizations of scattering potential in the sample.
Averaging Eq.4 over realizations of the random potential and using the 
approximation $\langle\Delta(\vec{r}) K(\vec{r},\vec{r'})\rangle=
\langle\Delta(\vec{r}; H)\rangle \langle K(\vec{r},\vec{r'}, H)\rangle$ we
get the above 
mentioned expression for $H^{0}_{c}$.
In the case of strong magnetic fields, when 
$\Delta(\vec{r}; H) \ll \Delta_0$, we can 
expand Eq.4 in terms of $\Delta(\vec{r})$. Since $\Delta(\vec{r})$ varies 
slowly over distances of the order of $\xi_0$,
while $\langle K(\vec{r},\vec{r'})\rangle$ decays 
exponentially for
$|\vec{r}-\vec{r'}|\gg \xi_0$, we can also make the gradient expansion of 
Eq.4. As a result we get from Eq.4 
\begin{eqnarray}
\left(\frac{1}{12}\xi_{0}^{2}(\nabla-\frac{2e}{c}\vec{A})^2 + \frac{H^0_c
-H}{H^0_c}\right)\Delta(\vec{r})+ 
\int \delta K^0(\vec r, \vec{r'},H,\vec{A})\Delta(\vec{r'}) d\vec{r'}
=o(\frac{\Delta^3(\vec r)}{\Delta_0}),
\end{eqnarray}
where $\delta K^0(\vec r, \vec{r'})=K^0(\vec r, \vec{r'}) - 
\langle K^0(\vec r, \vec{r'})\rangle$ and
$K^{0}(\vec{r},\vec{r'})=
K(\vec{r},\vec{r'}, \{\Delta(\vec r)=0\})$.  The 
difference between Eq.5 and the conventional Ginsburg-Landau equation
is the third term in Eq.5 which accounts for mesoscopic fluctuations
of the kernel $K^{0}(\vec{r},\vec{r'})$. It is precisely this term, which
at high magnetic fields leads to the random sign of superfluid density.

Employing the perturbation theory with respect to $\delta 
K^{0}(\vec{r},\vec{r'})$ we get from Eq.5 an expression for
the correlation function $C(\vec r_1-\vec r_2)=\langle\delta 
\Delta(\vec{r_1})
\delta \Delta(\vec{r_2})\rangle$
of the mesoscopic fluctuations of the superconducting order parameter
$\delta\Delta(\vec r; H)=\Delta(\vec r;H)-\langle\Delta(H)\rangle$

\begin{eqnarray}
C(\vec{\vec r_1}-\vec r_2)\approx
\frac{\Delta^2_0}{G^{2}}
\left \{ \begin{array}{cc}
1- (|\vec{r_1}-\vec r_2|/\xi(H))^2, & \mbox{$|\vec{r_1}-\vec r_2| \ll 
\xi(H)$}; \\
\exp(-2\sqrt{3}|\vec{r_1}-\vec r_2|/\xi(H)), & 
\mbox{$|\vec{r_1}-\vec r_2| \gg \xi(H)$}.
\end{array}\right.
\end{eqnarray}
In order to derive Eq.6 we had to calculate the correlation function 
$\langle\delta K^{0}(\vec{r_1},\vec{r_4}) \delta
K^{0}(\vec{r_2},\vec{r_3})\rangle$ using the diagrams shown in Fig.2b.  It
follows from Eq.6 that $\langle(\delta\Delta)^{2}\rangle$ in the
two-dimensional
case is almost independent of $H$, but $\langle\Delta(H)\rangle$
decreases with $H$. As a result, perturbation theory holds
as long as
${\langle\Delta(H)\rangle}/{\Delta_0}=\sqrt{(H^{0}_{c}-H)/H^{0}_{c}}\gg
G^{-1}$.

Using the expression for the supercurrent expanded in terms of
$\Delta(\vec{r},H)\ll \Delta_{0}$
we have for the correlation function
of the nonlocal superfluid density
$N_{s}(\vec{r},\vec{r'})$,
\begin{eqnarray}
\vec J_s(\vec r)=\frac{e^2}{m}\int d\vec{r'}N_s(\vec{r}-\vec{r'})
\vec{V}_{s}(\vec{r'})
=\frac{e^2}{m}\int d\vec{r'}(N_s(H)\delta(\vec{r}-\vec{r'})
+\delta N_s(\vec r, \vec{r'}))\vec{V}_{s}(\vec{r'}),
\nonumber \\
\frac{\langle\delta N_s(\vec r_1, \vec{r_1'}) \delta N_s(\vec r_2,
\vec{r_2'})\rangle}
{(N_{s}(H))^2}
\sim \frac{C^2(\vec r_1 -\vec r_2)}{\langle\Delta(H)\rangle^{4}}
\delta(\vec{r}_1 - \vec{r_1'})
\delta(\vec{r}_{2} - \vec{r_2'})+ 
\frac{G^{-2}\xi_0^4}{|\vec{r}_{1} -\vec{r_1'}|^4}
\delta(\vec{r}_{1} - \vec{r}_{2})
\delta (\vec{r_1'} - \vec{r_2'})
\end{eqnarray}
which is valid as long as 
${\langle\Delta(H)\rangle}/{\Delta_{0}}\gg G^{-1}$
and $\xi(H) \gg |\vec r_1-\vec r_2| \gg \xi_0$.
At $|\vec r_1-\vec r_2| \gg \xi(H)$, the correlation function
in Eq.7 becomes exponentially small.
Here $N_{s}(H)=N_s^0 {\langle\Delta(H)\rangle^{2}}/{\Delta_0^2}$, 
$N_{s}^{0}=N({l}/{\xi_0})^2$ is the average superfluid density
at $H=0$ and $N$ is the electron concentration in the metal.
The first term of the correlation function in Eq.7 is connected to the fluctuations of the order 
parameter $\Delta(\vec r)$ as shown in Fig.2c.
The second term corresponds to the fluctuations of the Green
functions $G_{\epsilon}(\vec{r},\vec{r}')$ shown in Fig.2d.

Therefore if the magnetic field is close to the critical one, i.e.
$|H-H_{c}^{0}|/H^{0}_{c}\sim G^{-2}$, then the amplitude of fluctuations
of the superfluid 
density averaged over the size $\xi({H})$ becomes of order of its the
average {$\delta N_{s}\sim \langle 
N_{s}\rangle$, which means that the 
local value of the superfluid density, averaged over the size $\xi_{0}$, 
becomes of random sign. Hence the system is unstable with
respect to spontaneous creation of supercurrents.

If ${|H-H^0_{c}|}/{H^0_{c}} \ll G^{-2}$, one can neglect the second
term in 
brackets in Eq.5. 
Rescaling $\vec{r}\sim\vec{x} G\xi_{0}, 
\Delta(\vec r)\sim{\Delta_0}/{G}f({\vec r}/{
G\xi_0})$, yields a dimensionless stochastic equation for $f(\vec{x})$

\begin{equation}
\nabla_{\vec{x}}^{2} f(\vec{x}) +\int d\vec{x'} \delta k(\vec{x},
\vec{x'}) f(\vec{x'})=f^3(\vec{x}), 
\end{equation}
where $\langle\delta k(\vec{x},\vec{x'})\rangle=0$ and the correlation
function
$\langle\delta k(\vec{x}_1, \vec{x'_1}) \delta k(\vec{x}_2,
\vec{x'_2})\rangle
=(\delta(\vec{x}_1 - \vec{x'_1})+ 
G^{-2}/\{G^{-4} +(\vec{x}_1 - \vec{x'_1})^4\})
\delta (\vec{x}_1 -\vec{x}_2)\delta
(\vec{x'_1}-\vec{x'_2})$ is given by diagrams shown in Fig.2b. 
It follows from Eq.8 that the amplitude 
of fluctuation of the 
modulus of the order parameter $\delta\Delta(\vec{r})\sim
\langle\Delta(H)\rangle \sim \Delta_0/ G$ is of order of 
its average. The characteristic spatial scale of the fluctuations of
$\delta\Delta(\vec{r})$ is of order of $L_D$.
The sign of second term in Eq.8
fluctuates randomly which corresponds to the random sign of the superfluid
density.
The spontaneously created supercurrents in this case have random
directions, their typical amplitude is
of order of $J^{c}_{s}\sim N_{s}^{0}{\hbar}/{G^3\xi_0}$ and their
characteristic scale of spatial correlations is also of order of $L_D$.
The current described by Eq.3 is negligible compared with $J^{c}_{s}$ when
$l \ll G^{-1}\sqrt{D\tau_{so}}$. 

The fact that the sign of $N_{s}$ is random is
especially 
clear in the case of large magnetic field, when
$H-H^0_c \gg H^0_c G^{-2}$. 
In this case, $\Delta(\vec r)$ is nonzero only due to existence of the rare
regions, where $\delta k(\vec{x},\vec{x'})$ is much larger than the
typical value. Thus, the spatial dependence of the modulus of the order 
parameter has the form of superconducting domains embedded in a normal
metal. These regions are 
connected via the Josephson effect. We can calculate the average critical 
current of the junctions and its variance as functions of the distance
between the droplets $L_{0}$:
\begin{eqnarray}
\langle J_{c}\rangle \sim
G\frac{e}{\hbar}\frac{D}{L_0^2}
\exp\left(-\frac{2L_0}{\sqrt{3}\xi(H)}\right); 
\langle(\delta
J_{c})^{2}\rangle\sim
\left(\frac{e}{\hbar}\frac{D}{L^2_0}\right)^2. 
\end{eqnarray}
They decay with $L_{0}$ exponentially and as a power law
respectively.
 As a result, the amplitude 
of the fluctuations turns out to be larger than the
average, which means that $J_{c}$ has a random sign. 

It is well known$^{[21]}$ that at $T=0$ the long range order of the
ground state of the two-dimensional $XY$ model is destroyed by an 
arbitrary small concentration of {\em antiferromagnetic}  bonds.
As we have mentioned above in the case $H\ll H^{0}_{c}$ regions, where
$N_{s}(\vec{r})<0$,
exist with small but finite probability. 
In this case, however, the properties of superconducting system are
different from the $XY$ model because the supercurrents
spontaneously created in these regions are screened
by the Meisner effect. Thus at $H,T=0$ 
superconducting films should exhibit the conventional long 
range order. This implies that there is a critical magnetic field 
$H_{SG}<H_{c}^{0}$ where at $T=0$
the system has a phase transition from superconducting to the
superconducting 
glass states $(H_{c}^{0}- H_{SG}\sim H_{c}^{0}G^{-2})$. The interval of
magnetic fields where the system is in the 
superconducting glass state is indicated by shaded region in Fig.1.

 
Let us now consider the case of weak spin-orbit 
scattering limit $\Delta_{0}\tau_{so}\gg 1$. 
In this case the spin magnetization in the superconducting phase is zero. 
Correspondingly, the 
conventional theory based on the equation for average order parameter leads
to the conclusion that 
the superconductor-normal metal transition is of first order 
with the critical magnetic field $H_{cc}$$^{[3,4]}$. 
However, the fluctuations of both 
magnetization energy of the normal metal and the condensation energy of
the superconductor phase should lead to a nonuniform state, 
qualitatively similar to the case $\Delta_{0}\tau_{so}\ll 1$.
The theory of this phenomenon at $\Delta_{0}\tau_{so}\gg 1$ is, however,
more difficult.  In this case a domains of normal phase within a bulk 
superconductor (or a superconducting domain in normal metal) has the
surface 
energy of order of $d L_{D}\xi_{0}\nu_0\Delta_0^2$, where 
$L_{D}$ is the domain size. This energy is much larger than 
the typical energy associated with mesoscopic fluctuations in Eq.2,
$d L_D \xi_0 \nu_0 (\mu_B H)^2G^{-1}$. 
Thus the probability of 
the occurrence of such domains is small as long as $G>1$. We would like to 
stress though, that qualitatively the case 
$\Delta_{0}\tau_{so}\gg 1$ is not 
different from the case $\Delta_{0}\tau_{so}\ll 1$ for in both cases the
 superconducting glass solutions survive at $T=0$ and $H>H^{0}_{c}$.    

The question whether or not the quantum fluctuations of the phase of the 
order parameter
destroy the superconducting glass state at $T=0$ and large $H$
is still open$^{[22-24]}$. 
At finite temperatures $T>0$, 
strictly speaking, 
the system considered above doesn't posses a phase rigidity
because of Meisner screening effect$^{[25]}$.
On the other hand the two dimensional $XY$ model with random sign of
exchange 
interaction is known not to exhibit a phase transition between the 
paramagnetic and the spin-glass phases$^{[26]}$, again implying the
absence of long range order.
We acknowledge useful discussions with B.Altshuler and S.Kivelson.
This work was supported by the Division of Material Sciences, U.S.National
Science Foundation under Contract No.DMR-9625370 and the US-Israeli
Binational Science Foundation grant no. 94-00243.


{\begin{figure}[h]
\begin{center}
\leavevmode
\epsfbox{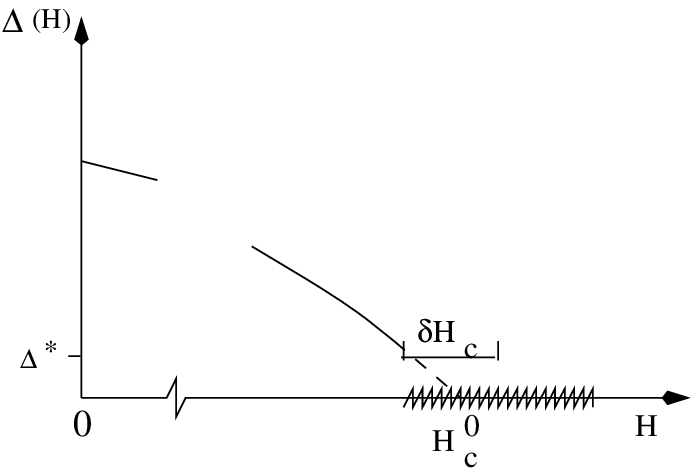}
\end{center}
\caption{
Qualitative picture of the magnetic field dependence of $\Delta(H)$ at
zero temperature when $\Delta_0\tau_{s0} \ll 1$.
The shaded region corresponds to the superconducting glass phase. 
$\Delta^* \sim \Delta_0/G$, $\delta H_{c} \sim H^0_{c}/G^2$.} 
\label{fig:1}
\end{figure}}

{
\begin{figure}[h] 
\begin{center}
\leavevmode
\epsfbox{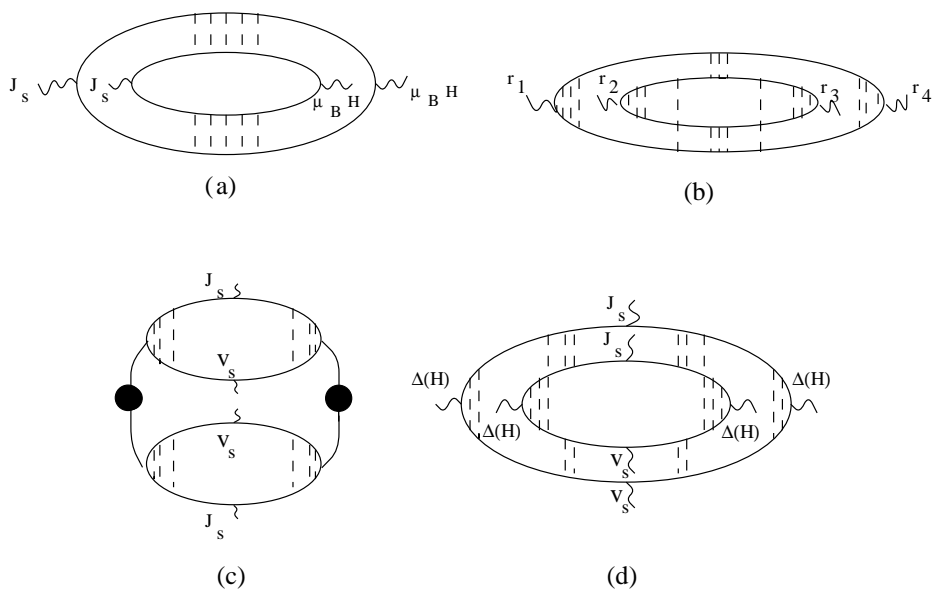}
\end{center}
\caption{
a) Diagrams representing the current correlation function Eq.3.
b) Diagrams representing the correlation function $\langle\delta
K^{0}(\vec{r_1},\vec{r_4})\delta
K^{0}(\vec{r_2},\vec{r_3})\rangle$. 
c)d) Diagrams representing the correlation function of
supercurrent densities
$\langle\vec{J}_{s}(\vec{r})\vec{J}_{s}(\vec{r'})\rangle$.   
Solid lines correspond to
electron Green functions in metal and dashed lines correspond to
elastic scatterings of a random potential and black dots represent the
correlation function $\langle\delta \Delta(\vec r_1) 
\delta \Delta(\vec r_2)\rangle$ given by Eq. 6.
}
\label{fig:2}
\end{figure}}

\end{document}